\documentclass[aip,amsmath,amssymb,reprint,pop]{revtex4-1}

\usepackage{graphicx}  
\usepackage{dcolumn}
\usepackage{bm}
\usepackage{amsfonts} 
\usepackage[utf8]{inputenc}
\usepackage[T1]{fontenc}
\usepackage{mathptmx}
\usepackage{etoolbox}

\def\E{\mbox{e}}
\def\dd{\mbox{d}}

\def\dt{\mbox{d}t}

\def\d3r{\mbox{d}^3r}

\def\e0{\varepsilon_0}

\def\beq{\begin{equation}}
\def\eeq{\end{equation}}
\def\bea{\begin{eqnarray}}
\def\eea{\end{eqnarray}}
\def\bi{\begin{itemize}}
\def\ei{\end{itemize}}
\def\bp{\begin{picture}}
\def\ep{\end{picture}}

\def\epsv{\mbox{\boldmath$\epsilon$}}

\def\Piv{\mbox{\boldmath$\Pi$}}

\def\ddt{\partial_t}

\def\dd{\mbox{$\mathrm d$}}

\def\Av{{\bf A}}
\def\Ev{{\bf E}}

\def\pv{{\bf p}}

\usepackage{color}

\makeatletter
\def\@email#1#2{%
 \endgroup
 \patchcmd{\titleblock@produce}
  {\frontmatter@RRAPformat}
  {\frontmatter@RRAPformat{\produce@RRAP{*#1\href{mailto:#2}{#2}}}\frontmatter@RRAPformat}
  {}{}
}%
\makeatother
\begin{document}

\title[]{Comment on: ``Interacting quantum and classical waves: Resonant and non-resonant energy transfer to electrons immersed in an intense electromagnetic wave'' [Phys. Plasmas 29, 022107 (2022)]}

\author{A. Macchi}%
\altaffiliation[Also at ]{Enrico Fermi Department of Physics, University of Pisa, Pisa, Italy}
\affiliation{National Institute of Optics, National Research Council (CNR/INO), Adriano Gozzini laboratory, Pisa, Italy}
\email{andrea.macchi@ino.cnr.it}

\date{\today}

\begin{abstract}
\end{abstract}

\maketitle

In Ref.\onlinecite{deception}, a theory of the interaction of a particle with an electromagnetic (EM) wave in a plasma based on the quantum Klein-Gordon equation was presented. In this Comment we show that the basic results of Ref.\onlinecite{deception} are recovered by a simple classical calculation, and we add remarks on the (un)suitability of the proposed mechanism for particle energization in both laser and astrophysical plasmas. 

Following Ref.\onlinecite{deception}, we consider in a reference frame $S$ a transverse, monochromatic wave in a tenuous plasma of density $n_0$ represented by the vector potential
\bea
\Av_{\perp}=\Av_{\perp}(\omega t-kz)
=A_0\mbox{Re}(\epsv\E^{-ikz+i\omega t}) \; ,
\label{eq:wave}
\eea
corresponding to a four-potential $A_{\mu}=(0,\Av_{\perp},0)$. The dispersion relation is
\bea
\omega^2=\omega_p^2+k^2c^2 \; ,
\eea
where $\omega_p=(4\pi e^2n_0/m_e)^{1/2}$ is the plasma frequency and $\omega>\omega_p$ is assumed for propagation. 
Corrections due to both thermal and relativistic dynamics effects may be incorporated in a redefinition of $\omega_p$, see Eq.(33) in Ref.\onlinecite{deception}.

Let us consider the interaction of the wave (\ref{eq:wave}) with a particle of charge $q$ and mass $m$ which at $(z,t)=(0,t_i)$ has a 4-momentum
\bea
p_{i}&=&(p_{i0},\pv_i)=(p_{i0},\pv_{i\perp},p_{iz}) \, , \\ 
p_{i0}&=&(m^2c^2+\pv_i^2)^{1/2} \; .
\eea
The initial conditions are fully general. Now it is advantageous to switch to a Lorentz frame $S'$ moving in the $z$-direction at the group velocity $v_g=\partial\omega/\partial k$ with boost parameters
\bea
\beta_g=\frac{v_g}{c}=\frac{kc^2}{\omega} \; , \qquad
\gamma_g=(1-\beta_g^2)^{-1/2}=\frac{\omega}{\omega_p} \; .
\label{eq:boost}
\eea
In $S'$, the vector potential depends only on time $t'$ and there is no magnetic field, as it is easily found from the Lorentz transformations of the wave phase and the four-potential:
\bea
\omega t -kz= \omega_p t' \; , \qquad
A'_{\mu}=(0,\Av_{\perp}(\omega_p t'),0) \; .
\eea
The solution to the equation of motion for $t'>t'_i=\gamma_g t_i$ 
is thus straightforward. From
\bea
\frac{\dd \pv'_{\perp}}{\dt'}=q\Ev'(t')=-\frac{q}{c}\ddt'\Av_{\perp}\; , \qquad
\frac{\dd p'_z}{\dt'}=0 \; ,
\eea
one obtains $p'_z(t')=p'_{iz}=\gamma_g(p_{iz}-\beta_g p_{i0})$ and
\bea
\pv_{\perp}'(t')+\frac{q}{c}\Av(\omega_p t')=\pv'_{i\perp}+\frac{q}{c}\Av_i \equiv \Piv_i \; ,
   \label{eq:Piv}
\eea
where $\Av_i\equiv \Av_{\perp}(\omega_p t'_i)$; note that $\pv'_{i\perp}=\pv_{i\perp}$. Eq.(\ref{eq:Piv}) gives the conservation of the canonical momentum $\Piv=\pv_{\perp}+\Av_{\perp}$. The 4-momentum in the boosted frame is
\bea
p'_{\mu}=\left(p'_{0},\pv'_{\perp},p'_{iz}\right)
\, , 
\eea
where
\bea
p'_0=\left(\pv^{'2}_{\perp}+p^{'2}_{iz}+m^2c^2\right)^{1/2} \; .
\eea
By transforming back in the $S$ frame we obtain for the energy
\bea
p_{0}&=&\gamma_g(p'_{0}+\beta_g p'_{iz}) \\
&=& \gamma_g(p'_{0}+\beta_g\gamma_g(p_{iz}-\beta_g p_{i0})) \; .
\label{eq:p0}
\eea
In Ref.\onlinecite{deception} it is stated that energization  occurs when the initial particle velocity along $v_z=p_zc/p_0=v_g$, the group velocity. Actually Ref.\onlinecite{deception} reports ``the phase velocity of the EM wave equals to the axial (in the propagation direction) phase velocity of the relativistic quantum particle'' which is an obvious mistake (presumably a misprint) since the EM phase velocity $v_p=\omega/k>c$, so there cannot ever be ``phase matching'' between the particle and the wave. Looking at Eq.(\ref{eq:p0}) we see instead that $p_{iz}\geq \beta_g p_{i0}$ is necessary to make  the second term in Eq.(\ref{eq:p0}) positive. When $p_{iz}=\beta_g p_{i0}$ (corresponding to the erroneous ``resonant'' condition of Ref.\onlinecite{deception}) the particle velocity along $z$ is null in $S'$ ($p'_z=p'_{iz}=0$) and constant $(v_z=v_g)$ in $S$ (note that this does not imply $\dd p_z/\dt =0$ in $S$; it is a consequence that, being $B/E=\beta_g$, $\dd p_z/\dt =\beta_g\dd p_0/\dt$), and the second term in the time-dependent energy (\ref{eq:p0}) vanishes. Thus we obtain
\bea
p_{0}(t)&=&\frac{\omega}{\omega_p}\left(\left(\Piv_i-\frac{q}{c}\Av_{\perp}(\xi)\right)^2+m^2c^2\right)^{1/2} \nonumber\\
       &=&\left(p^2_{i0}+\left(\frac{\omega}{\omega_p}\right)^2
\left(\frac{q^2}{c^2}(\Av^2_{\perp}(\xi)+\Av^2_i) \right.\right. \nonumber \\ & & \left. \left. -2\frac{q}{c}\left(A_{\perp}(\xi)\Pi_i-A_ip_{i\perp}\right)\right)\right)^{1/2}
\label{eq:pf0}
\eea
where
\bea
\xi=\omega t -k(v_g t)=\omega t (1-\beta_g^2)=\frac{\omega_p^2}{\omega}t \; ,
\eea
and we used Eqs.(\ref{eq:Piv}) and (\ref{eq:boost}).

{For the case $\Piv_i=0$, Eq.(\ref{eq:pf0}) is identical to Eq.(33) of Ref.\onlinecite{deception}, namely Eq.(33) for the case $\Piv_i=0$. For $\Piv_i\neq 0$, Eq.(44) of Ref.\onlinecite{deception} yields $p_0=\left\langle{\cal E}\right\rangle/c$ where
  \bea
  \left\langle{\cal E}\right\rangle =\frac{\omega}{\omega_p}\left(\Piv_i^2+\left(\frac{q}{c}A_{\perp}\right)^2+m^2c^2\right)^{1/2} \, ,
  \eea
  after recognizing that in such equations $K_{\perp}c$ is the conserved canonical momentum $\Pi_i$. The difference with our Eq.(\ref{eq:pf0}) comes from the mistake in Ref.\onlinecite{deception} of calculating the energy as the expectation value
  \bea
  \left\langle{\cal E}\right\rangle=\left\langle i\hbar\Psi^*\partial_t\Psi\right\rangle \, ,
  \eea
  with $\psi$ the wavefunction. However, as it fully evident in Schr\"odinger equation (2) of Ref.\onlinecite{deception}, the operator $i\hbar\ddt$ does not correspond to the particle energy $\pv^2/2m$ but rather to $(\pv+q\Av/c)^2/2m=(p_z+\Piv)^2/2m$. The same mistake occurs for the Klein-Gordon equation where the coupling with the classical EM field is also made by the replacement $\pv \to \pv+q\Av/c$. Hence in the results of Ref.\onlinecite{deception} the canonical momentum $K_{\perp}c=\Pi_i$ should be replaced by the transverse kinetic momentum $\pv_{\perp}$. Looking at the first expression of Eq.(\ref{eq:pf0}) we see that $\Piv_i-({q}/{c})\Av_{\perp}(\xi)=\pv_{\perp}$, thus (\ref{eq:pf0}) is identical to the quantum result when the expectation value of the energy is calculated correctly.
}

Recovering the results of Ref.\onlinecite{deception} via a classical calculation shows that a quantum approach is not necessary. This could be expected \emph{a priori} since the de Broglie length of high-momentum particle is very small, so that the particle behaves classically.
{The statement in Ref.\onlinecite{deception} ``the wave–particle coupling is fully contained in the term proportional to $K_{\perp}$'' [i.e the term containing $A_{\perp}\Piv_i$ in Eq.(\ref{eq:pf0})] might suggest that such term is of quantum nature, but indeed it has a simple classical origin in the  ${\bf v}\times{\bf B}$ force: when $\Piv_i\neq 0$ the particle has an initial velocity in the perpendicular direction which yields a force term $\propto p_{\perp}B/p_0$.}

{
  The ``rate of change of particle energy'' can be trivially and explicitly calculated as $\dd p_0/\dt$ from Eq.(\ref{eq:pf0}). However, Ref.\onlinecite{deception} rather considers the root mean square of $\left\langle{\cal E}\right\rangle$,
  \bea
  \frac{\dd}{\dt}\left\langle{\cal E}\right\rangle \propto \left(\frac{\omega}{\omega_p}\right)^2\frac{A_{\perp}\Pi_{\perp}}{\left\langle{\cal E}\right\rangle}\frac{\sin kL}{kL} \; ,
  \label{eq:rms}
  \eea
where the factor ${\sin kL}/{kL}$ originates from the particle being now assumed to be localized in a region of length $L$ [Eq.(E5) of Ref.\onlinecite{deception}]. While admittedly the need for and meaning of (\ref{eq:rms}) are not fully clear, because of its definition (\ref{eq:rms}) should represent a sort of uncertainty in the value of ${\dd}{\cal E}/{\dt}$; actually because of Heisenberg's principle the uncertainty in the longitudinal momentum should be proportional to $1/L$. Moreover, the values of the energy may have zero uncertainty only if the particle is in a steady state after ``localization'', which is possible only for ``standing'' wavefunctions with $\sin kL=0$ so that the region contains an integer number of half de Broglie wavelengths. This would be the only truly quantum effect described in Ref.\onlinecite{deception}. Clearly, for the scenarios discussed below $kL$ has enormous values so that ${\dd}{\cal E}/{\dt}\to 0$; quantum uncertainty effects are completely negligible.
}

The conceptual transparency of the classical description provides insight on the possibility of transverse EM waves in a low-density plasma to yield ultra-high energization of particles. First of all, modeling the EM field as a plane, monochromatic wave of infinite width and duration is at best a local approximation for an EM field of finite extension in space and time. Assuming an adiabatic rise and fall of the field from $A_i=0$ to its peak value and than back to zero, no energy is eventually left to the particle {(this is similar to the ``no acceleration'' or Lawson-Woodward theorem for a plane wave in vacuum\cite{gibbon,macchi})}. This is why we write ``energization'' instead of ``acceleration'' which would suggest that the ultra-high energy particles can still be detected out of the interaction region.
{A net energy only appears for $A_i\neq 0$ which may occur for a particle released from rest when the EM field is already on (a field-ionized electron, for example).}

If a quasi-monochromatic wavepacket is considered, the initial condition $v_{iz}=v_g$ requires the particle to be ``born'' inside the wavepacket which otherwise would never be reached. Even if the particle is efficiently injected (by some unspecified mechanism) into the wavepacket with $v_{iz}=v_g$, reaching the maximum possible energy gain would require in $S'$ a time $t'_m=\pi/(2\omega_p)$  (equal to a quarter of the oscillation period) which, due to Lorentz time dilation, corresponds to $t_m=\gamma_gt'_m=(\pi\omega/2\omega_p^2)$ in the laboratory frame $S$ as also indicated by the temporal dependence of $A$ in Eqs.(\ref{eq:pf0}). This greatly increased energization time may be not compatible with the assumption of a quasi-plane, quasi-monochromatic field. Picking up an example from Ref.\onlinecite{deception}, we observe that using high-intensity femtosecond laser pulses laser-plasma interaction experiments are nowadays routinely performed with parameters such as $a_0=eA/mc^2=10$ and $\omega/\omega_p=10^2$. An energy gain factor of $\sim a_0(\omega/\omega_p)$ would require an interaction time of $\sim (\omega/\omega_p)^2=10^4$ laser cycles which is typically $\sim 10^2$ times the pulse duration, and would imply particles to move over $\sim$mm distances which are much larger than both the Rayleigh length and focal width of the laser beam. These numbers largely justify the lack of any evidence of ultrahigh energization in laser-plasma experiments. For what concerns an astrophysical scenario with $\omega/\omega_p=10^5$, assuming a frequency of $\omega/2\pi=10^{11}$~Hz the EM field should keep a constant amplitude over a distance of $\sim 5 \times 10^6$~km in both transverse and longitudinal directions.

\end{document}